\documentstyle[12pt,epsfig]{article}
\def\cl{\centerline}
\def\vs{\vskip}

\def\be{\begin{equation}}
\def\ee{\end{equation}}
\def\bea{\begin{eqnarray}}
\def\eea{\end{eqnarray}}

\begin{document}
\hfill{NCKU-HEP-99-03}\par
\vs 0.3cm
\begin{center}
{\large {\bf Soft Gluons in Logarithmic Summations} }
\vs 1.0cm
Hsiang-nan Li
\vs 0.3cm
Department of Physics, National Cheng-Kung University, \par
Tainan 701, Taiwan, Republic of China
\vs 1.0cm
Jyh-Liong Lim
\vs 0.3cm
Department of ElectroPhysics, National Chiao-Tung University, \par
Hsinchu 300, Taiwan, Republic of China
\end{center}

\vs 1.0cm

\cl{\bf Abstract}
\vs 0.3cm
We demonstrate that all the known single- and double-logarithm summations 
for a parton distribution function can be unified in the Collins-Soper
resummation technique by applying soft approximations appropriate in
different kinematic regions to real gluon emissions. Neglecting the gluon
longitudinal momentum, we obtain the $k_T$ (double-logarithm) resummation
for two-scale QCD processes, and the Balitsky-Fadin-Kuraev-Lipatov
(single-logarithm) equation for one-scale processes. Neglecting the
transverse momentum, we obtain the threshold (double-logarithm) resummation
for two-scale processes, and the Dokshitzer-Gribov-Lipatov-Altarelli-Parisi
(single-logarithm) equation for one-scale processes. If keeping the
longitudinal and transverse momenta simultaneously, we derive a unified
resummation for large Bjorken variable $x$, and a unified evolution equation
appropriate for both intermediate and small $x$.

\newpage

\cl{\large \bf 1. Introduction}
\vs 0.5cm

It is known that radiative corrections in perturbative QCD (PQCD) produce
large logarithms at each order of the coupling contant. Double logarithms
appear in processes involving two scales, such as $\ln^2(p^+b)$ with
$p^+$ the large longitudinal momentum of a parton and $1/b$ the small
inverse impact parameter, where $b$ is conjugate to the parton transverse
momentum $k_T$. In the kinematic end-point region with large Bjorken
variable $x$, there exist $\ln^2(1/N)$ from the Mellin transformation of
$\ln(1-x)/(1-x)_+$, for which the two scales are the large $p^+$ and the
small infrared cutoff $(1-x)p^+$ for gluon emissions from the parton.
Single logarithms are generated in processes involving only one scale, such
as $\ln p^+$ and $\ln(1/x)$, for which the relevant scales are the large
$p^+$ and the small $xp^+$, respectively. These logarithmic corrections
to a parton distribution function have been summed to all orders by various
methods, which are the $k_T$ resummation for $\ln^2(p^+b)$ \cite{CS}, the
threshold resummation for $\ln^2(1/N)$ \cite{S,CT,KM}, the
Dokshitzer-Gribov-Lipatov-Altarelli-Parisi (DGLAP) equation for $\ln p^+$
\cite{AP}, and the Balitsky-Fadin-Kuraev-Lipatov (BFKL) equation for
$\ln(1/x)$ \cite{BFKL}.

In this paper we shall demonstrate that all the above single- and
double-logarithm summations can be derived in the Collins-Soper (CS)
resummation technique \cite{CS}. The point is the soft approximation for
real gluon emissions, with which a parton distribution function
$\phi(x+l^+/p^+,|{\bf k_T+\bf l_T}|)$ is associated. The arguments of
$\phi$ indicate that the parton, emerging from a hadron, carries the
longitudinal momentum $xp^++l^+$ and the transverse momentum
${\bf k}_T+{\bf l}_T$ in order to radiate a real gluon with the momentum
$l$. If neglecting the $l^+$ dependence,
\begin{equation}
\phi(x+l^+/p^+,|{\bf k_T+\bf l_T}|)\approx
\phi(x,|{\bf k_T+\bf l_T}|)\;,
\label{nl}
\end{equation}
the scale $(1-x)p^+$ will not appear. Hence, Eq.~(\ref{nl}) is
inappropriate for the region with large $x\to 1$. Under this soft
approximation, we derive the $k_T$ resummation for intermediate $x$, which
involves two scales: the large $xp^+$ and the small $k_T$, and the BFKL
equation for small $x$, which involves only one scale $xp^+ \approx k_T$.
If neglecting the $l_T$ dependence,
\begin{equation}
\phi(x+l^+/p^+,|{\bf k_T+\bf l_T}|)\approx
\phi(x+l^+/p^+,k_T)\;,
\label{nk}
\end{equation}
the transverse degrees of freedom of a parton can be integrated out, and
$k_T$ will not be a relevant scale. Therefore, Eq.~(\ref{nk}) is
inappropriate for small $x$, where the scale $k_T$ is of order $xp^+$, and
not negligible. Under this soft approximation, we derive the threshold
resummation for large $x$, which involves two scales: the large $xp^+$ and
the small $(1-x)p^+$, and the DGLAP equation for intermediate $x$, which
involves only one scale $xp^+\sim (1-x)p^+$.

In the regions where Eqs.~(\ref{nl}) and (\ref{nk}) are inappropriate, we
should keep both the $l^+$ and $l_T$ dependences, and employ
$\phi(x+l^+/p^+,|{\bf k_T+\bf l_T}|)$ for real gluon emissions
directly. In this case the three scales $xp^+$, $(1-x)p^+$ and $k_T$ exist
simultaneously. We shall derive a unified resummation (a unification of the
$k_T$ and threshold resummations) for large $x$, and a unified evolution
equation (a unification of the DGLAP and BFKL equations), which is suitable
for both intermediate and small $x$. In conclusion, we are able to reproduce
all the logarithmic summations and derive their unifications simply by
employing appropriate soft approximations for real gluon emissions in the CS
technique. The results are summarized in Table I.

\vs 1.0cm
\cl {\large \bf 2. Master Equation}
\vs 0.5cm

Consider a parton distribution function $\phi(x,k_T,p^+)$ for a hadron with
the light-like momentum $p^\mu=p^+\delta^{\mu+}$, which describes the
probability that a parton carries the longitudinal momentum $xp^+$ and the
transverse momentum $k_T$. If the parton is a quark, $\phi$
is written, in the minimal subtraction scheme, as
\be
\phi(x,k_T,p^+)=\int\frac{dy^-}{2\pi}\int\frac{d^2y_T}{4\pi^2}
e^{-ix p^+y^-+i{\bf k}_T\cdot {\bf y}_T}
\langle p| {\bar q}(y^-,{\bf y}_T)\frac{1}{2}\gamma^+q(0)|p\rangle\;,
\label{deq}
\ee
where $\gamma^+$ is a Dirac matrix, and $|p\rangle$ denotes the hadron.
Averages over spin and color are understood. If the parton is a gluon, the
operator in the hadronic matrix element is replaced by
$F^+_\mu(y^-,{\bf y}_T)F^{\mu+}(0)$. The above definition is given in the
axial gauge $n\cdot A=0$ with the gauge vector $n^\mu=\delta^{\mu-}$ lying
on the light cone. To implement the CS technique, we allow $n$ to vary
arbitrarily away from the light cone ($n^2\not= 0$) \cite{CS}, and the
parton distribution function becomes gauge dependent. However, it will be
observed that the kernels for various logarithmic summations turn out to be
$n$-independent. This is natural, since it has been shown that parton
distribution functions defined for different $n$ possess the same infrared
structure, and thus the same evolutoin behavior, though different
ultraviolet structure \cite{L5}. After the derivation, we bring $n$ back to
the light cone, and the gauge invariance of the parton distribution funciton
is restored. That is, the arbitrary vector $n$ appears only at the
intermediate stage of the derivation, and acts as an auxiliary tool.

The master equation in the CS technique is a differential equation of
$\phi$ in $p^+$ \cite{CS,L1}. Because of the scale invariance of $\phi$ in
$n$ as indicated by the gluon propagator, $-iN^{\mu\nu}(l)/l^2$, with
\be
N^{\mu\nu}=g^{\mu\nu}-\frac{n^\mu l^\nu+n^\nu l^\mu}
{n\cdot l}+n^2\frac{l^\mu l^\nu}{(n\cdot l)^2}\;,
\label{gpp}
\ee
$\phi$ depends on $p^+$ via the ratio $(p\cdot n)^2/n^2$. Hence, we have
the chain rule relating the derivative $d\phi/dp^+$ to $d\phi/dn_\alpha$,
\be
p^+\frac{d}{dp^+}\phi=-\frac{n^2}{v\cdot n}v_{\alpha}
\frac{d}{dn_\alpha}{\phi}\;,
\label{cr}
\ee
with $v$ a dimensionless vector along $p$. The operator $d/dn_\alpha$
applies to gluon propagators, leading to 
\be
\frac{d}{dn_\alpha}N^{\mu\nu}=
-\frac{1}{n\cdot l}(l^\mu N^{\alpha\nu}+l^\nu N^{\mu\alpha})\;.
\label{dgp}
\ee
The loop momentum $l^\mu$ ($l^\nu$) contracts with the vertex the
differentiated gluon attaches, which is then replaced by a special vertex,
\bea
{\hat v}_\alpha=\frac{n^2v_\alpha}{v\cdot nn\cdot l}\;.
\label{va}
\eea
This special vertex can be read off from the combination of Eqs.~(\ref{cr})
and (\ref{dgp}).

Employing Ward identities \cite{L0}, a diagram with the contraction of
$l^\mu$ can be expressed as the difference of the diagram, in which
the particle (quark or gluon) propagator after the contraction is removed,
and the diagram, in which the particle propagator before the contraction
is removed. Hence, a pair cancellation exists between the diagrams with
adjacent contractions of $l^\mu$. The summation of all the diagrams with
different differentiated gluons then reduces to a single new diagram, where
the most external particle propagator is removed. That is, the special
vertex appears at the outer end of the parton line in this new diagram.
We obtain the master equation \cite{CS,L1},
\be
p^+\frac{d}{dp^+}\phi(x,k_T,p^+)=2{\bar \phi}(x,k_T,p^+)\;,
\label{meq}
\ee
shown in Fig.~1(a), where the new diagram denoted by $\bar \phi$ contains
the special vertex represented by a square. The coefficient 2 comes
from the equality of $\bar\phi$ with the special vertex on
either of the two parton lines.


The collinear region of the loop momentum $l$ is not important because
of the factor $1/(n\cdot l)$ in ${\hat v}_\alpha$ with nonvanishing $n^2$.
Therefore, the important regions of $l$ are soft and hard, in which the
subdiagram containing the special vertex can be factorized from
${\bar \phi}$ according to Figs.~1(b) and 1(c) at lowest
order, respectively. The second subdiagram in Fig.~1(c), as a soft
subtraction, guarantees a hard momentum flow. The remaining part is the
original distribution function $\phi$. Therefore, ${\bar \phi}$ is
factorized into the convolution of the subdiagram containing the special
vertex with $\phi$.

The soft contribution from Fig.~1(b) is written as
\be
{\bar \phi}_s(x,k_T,p^+)={\bar \phi}_{sv}(x,k_T,p^+)+
{\bar \phi}_{sr}(x,k_T,p^+)\;,
\label{ssf}
\ee
with
\bea
{\bar \phi}_{sv}&=&\left[ig^2C_F\mu^\epsilon
\int\frac{d^{4-\epsilon}l}{(2\pi)^{4-\epsilon}}
N_{\nu\beta}(l)\frac{{\hat v}^\beta v^\nu}{v\cdot l l^2}
-\delta K\right]\phi(x,k_T,p^+)\;,
\label{fsv} \\
{\bar \phi}_{sr}&=&ig^2C_F\mu^\epsilon
\int\frac{d^{4-\epsilon}l}{(2\pi)^{4-\epsilon}}
N_{\nu\beta}(l)\frac{{\hat v}^\beta v^\nu}{v\cdot l}
2\pi i\delta(l^2)
\nonumber\\
& &\times \phi(x+l^+/p^+,|{\bf k}_T+{\bf l}_T|,p^+)\;,
\label{fsr}
\eea
corresponding to the virtual and real gluon emissions, respectively. The
color factor $C_F=4/3$ should be replaced by $N_c=3$ in the case with the
parton being a gluon. The additive counterterm $\delta K$ is specified in
the modified minimal subtraction scheme. The hard contribution from
Fig.~1(c) is given by
\be
{\bar\phi}_h(x,k_T,p^+)=G(xp^+/\mu,\alpha_s(\mu))\phi(x,k_T,p^+)\;,
\label{gd}
\ee
with the hard function
\begin{eqnarray}
G&=&-ig^2C_F\mu^\epsilon\int\frac{d^{4-\epsilon}l}{(2\pi)^{4-\epsilon}}
N_{\nu\beta}(l)\frac{{\hat v}^\beta}{l^2}
\left[\frac{x\not p-\not l}{(xp- l)^2}\gamma^\nu
+\frac{v^\nu}{v\cdot l}\right]
-\delta G\;,
\nonumber\\
&=&-\frac{\alpha_s(\mu)}{\pi}C_F\ln\frac{xp^+\nu}{\mu}
\;,
\label{gh}
\end{eqnarray}
where $\delta G$ is an additive counterterm. In the case with the parton
being a gluon, the expression of $G$ can be written down straightforwardly.
The gauge factor $\nu=\sqrt{(v\cdot n)^2/|n^2|}$ confirms our argument that
$\phi$ depends on $p^+$ via the ratio $(p\cdot n)^2/n^2$.

\vs 1.0cm
\cl{\large \bf 3. $k_T$ Resummation and BFKL Equation}
\vs 0.5cm

We first discuss the soft approximation in Eq.~(\ref{nl}) for $\phi$
associated with the real gluon emission. Fourier transforming
Eq.~(\ref{fsr}) into the impact parameter $b$ space in order to decouple
the $l_T$ integration, we derive
\be
{\bar\phi}_s(x,b,p^+)=K(1/(b\mu),\alpha_s(\mu))\phi(x,b,p^+)\;,
\ee
with the soft function
\bea
K&=&ig^2C_F\mu^\epsilon\int\frac{d^{4-\epsilon}l}{(2\pi)^{4-\epsilon}}
N_{\nu\beta}(l)\frac{{\hat v}^\beta v^\nu}{v\cdot l}
\left[\frac{1}{l^2}+2\pi i\delta(l^2)e^{i{\bf l}_T\cdot {\bf b}}\right]
-\delta K\;,
\nonumber\\
&=&\frac{\alpha_s(\mu)}{\pi}C_F \ln \frac{1}{b\mu}
\;.
\label{kh}
\eea
Hence, in the intermediate $x$ region
$\phi$ involves two scales, the large $xp^+$ that characterizes the hard
function $G$ in Eq.~(\ref{gh}) and the small $1/b$ that characterizes the
soft function $K$.

Using ${\bar\phi}={\bar\phi}_s+{\bar\phi}_h$, the master equation
(\ref{meq}) becomes
\be
p^+\frac{d}{dp^+}\phi(x,b,p^+)=2\left[K(1/(b\mu),\alpha_s(\mu))+
G(xp^+/\mu,\alpha_s(\mu))\right]\phi(x,b,p^+)\;.
\label{dph}
\ee
Since both the ultraviolet divergences in $K$ and $G$ come from the virtual
gluon contribution ${\bar \phi}_{sv}$, they cancel each other, such that
$K+G$ is renormalization-group (RG) invariant. The single logarithms
$\ln(b\mu)$ and $\ln(xp^+/\mu)$, contained in $K$ and $G$, respectively,
are organized by the RG equations
\be
\mu\frac{d}{d\mu}K=-\gamma_K=-\mu\frac{d}{d\mu}G\;.
\label{kg}
\ee
The anomalous dimension of $K$, $\lambda_K=\mu d\delta K/d\mu$,
is given, up to two loops, by \cite{BS}
\be
\gamma_K=\frac{\alpha_s}{\pi}C_F+\left(\frac{\alpha_s}{\pi}
\right)^2C_F\left[{C}_A\left(\frac{67}{36}
-\frac{\pi^{2}}{12}\right)-\frac{5}{18}n_{f}\right]\;,
\label{lk}
\ee
with $n_{f}$ the number of quark flavors and $C_A=3$ a color factor.
The solution of Eq.~(\ref{kg}) gives
\bea
K(1/(b\mu),\alpha_s(\mu))+G(xp^+/\mu,\alpha_s(\mu))=
-\int_{1/b}^{xp^+}\frac{d\mu}{\mu}\gamma_K(\alpha_s(\mu))\;,
\label{skg}
\eea
where the initial conditions $K(1,\alpha_s(1/b))$ and $G(1,\alpha_s(xp^+))$
that contribute only to the single-logarithm summation have been droped.
Solving the differential equation (\ref{dph}) with the above expression
inserted, we obtain the $k_T$ resummation \cite{L1},
\bea
\phi(x,b,p^+)=\Delta_k(b,xp^+)\phi^{(0)}(x)\;,
\label{sph}
\eea
with the (Sudakov) exponential
\be
\Delta_k(b,xp^+)=\exp\left[-2\int_{1/b}^{xp^+}\frac{d p}{p}
\int_{1/b}^{p}\frac{d\mu}{\mu}\gamma_{K}(\alpha_s(\mu))\right]\;.
\label{fb}
\ee

In the small $x$ region with $xp^+\sim k_T$, or $xp^+\sim 1/b$ in the
$b$ space, the above two-scale case reduces to the one-scale case. The
source of double logarithms, {\it i.e.}, the integral containing $\gamma_K$
in Eq.~(\ref{skg}), is less important. Instead of applying the RG
equation (\ref{kg}), we simply add Eqs.~(\ref{fsv})-(\ref{gd}), or
equivalently, Figs.~1(b) and 1(c). The result can be understood in the way
that the function $G$ introduces an ultraviolet cutoff of order
$xp^+\sim k_T$, which comes from the first subdiagram of Fig.~1(c), to the
virtual soft gluon contribution. Without Fourier transformation,
${\bar \phi}$ can be reexpressed, according to Eqs.~(\ref{fsv}) and
(\ref{fsr}), as
\begin{eqnarray}
{\bar \phi}(x,k_T,p^+)&=&
ig^2N_c\int\frac{d^{4}l}{(2\pi)^4}N_{\nu\beta}(l)
\frac{{\hat v}^\beta v^\nu}{v\cdot l}
\left[\frac{\theta(k_T^2-l_T^2)}{l^2}\phi(x,k_T,p^+)\right.
\nonumber \\
& &\left.+2\pi i\delta(l^2)\phi(x,|{\bf k}_T+{\bf l}_T|,p^+)\right]\;,
\label{kf1}
\end{eqnarray}
where the color factor has been replaced by $N_c$, because we consider the
gluon distribution function in the small $x$ region. The $\theta$ function,
defining the ultraviolet cutoff $k_T$, is the consequence of the
inclusion of $G$.

To make a variation in $x$ via a variation in $p^+$, we assume a fixed
parton momentum. Under this assumption, the momentum fraction $x$ is
proportional to $1/p^+$, and we have \cite{L0}
\begin{equation}
p^+\frac{d}{dp^+}\phi(x,k_T,p^+)=-x\frac{d}{dx}\phi(x,k_T,p^+)\;.
\label{vac}
\end{equation}
Performing the integrations over $l^+$ and $l^-$ in Eq.~(\ref{kf1}) and
using Eq.~(\ref{vac}), the master equation (\ref{meq})
reduces to the BFKL equation \cite{KMS},
\begin{eqnarray}
\frac{d\phi(x,k_T,p^+)}{d\ln(1/x)}=
{\bar \alpha}_s\int\frac{d^{2}l_T}{\pi l_T^2}
\left[\phi(x,|{\bf k}_T+{\bf l}_T|,p^+)
-\theta(k_T^2-l_T^2)\phi(x,k_T,p^+)\right],
\label{bfkl}
\end{eqnarray}
with the coupling constant ${\bar \alpha}_s=N_c\alpha_s/\pi$.

\vs 1.0cm
\cl{\large \bf 4. Threshold Resummation and DGLAP Equation}
\vs 0.5cm

We then consider the soft approximation in Eq.~(\ref{nk}). In this case
the dependence on $k_T$ can be integrated out from both sides of
Eqs.~(\ref{fsv})-(\ref{gd}), and the scale $(1-x)p^+$ enters. We employ the
Mellin transformation to bring ${\bar\phi}_{sr}$ from the momentum fraction
$x$ space to the moment $N$ space,
\bea
{\bar\phi}_{sr}(N,p^+)&=&\int_0^1 dxx^{N-1}{\bar\phi}_{sr}(x,p^+)\;,
\eea
under which the $l^+$ integration decouples. Combined with the soft virtual
contribution in Eq.~(\ref{fsv}), we derive 
\be
{\bar\phi}_s(N,p^+)=K(p^+/(N\mu),\alpha_s(\mu))\phi(N,p^+)\;,
\ee
with the soft function
\bea
K&=&ig^2C_F\mu^\epsilon\int_0^1dz
\int\frac{d^{4-\epsilon}l}{(2\pi)^{4-\epsilon}}
N_{\nu\beta}(l)\frac{{\hat v}^\beta v^\nu}{v\cdot l}
\left[\frac{\delta(1-z)}{l^2}\right.
\nonumber\\
& &\left.+2\pi i\delta(l^2)\delta\left(1-z-\frac{l^+}{p^+}\right)
z^{N-1}\right]-\delta K\;,
\nonumber\\
&=&\frac{\alpha_s(\mu)}{\pi}C_F\ln\frac{p^+\nu}{N\mu}\;,
\label{ktt}
\eea
and the counterterm $\delta K$ the same as that in Eq.~(\ref{kh}).
Therefore, in the large $x$ region $\phi$ involves two scales, the large
$xp^+\sim p^+$ from the hard function $G$ in Eq.~(\ref{gh}) and the small
$(1-x)p^+\sim p^+/N$ from the soft function $K$.

Similarly, Eqs.~(\ref{dph})-(\ref{skg}) hold but with $1/b$ replaced by
$p^+/N$. To sum $\ln(1/N)$, we regard the derivative $p^+d\phi/dp^+$ as
\begin{equation}
p^+\frac{d\phi}{dp^+}=
\frac{p^+}{N}\frac{\partial\phi}{\partial (p^+/N)}\;,
\end{equation}
which leads to the threshold resummation,
\begin{eqnarray}
\phi(N,p^+)=\Delta_t(N,p^+)\phi^{(0)}
\label{pht}
\end{eqnarray}
with the exponential
\begin{eqnarray}
\Delta_t(N,p^+)=\exp\left[-2\int_{p^+/N}^{p^+}\frac{d p}{p}
\int_{p}^{p^+}\frac{d\mu}{\mu}
\gamma_{K}(\alpha_s(\mu))\right]\;.
\label{fbt}
\end{eqnarray}

In the intermediate $x$ region the above two-scale case reduces to the
one-scale case because of $xp^+\sim (1-x)p^+$, and the source of double
logarithms becomes less important. Without the Mellin transformation,
the addition of Eqs.~(\ref{fsv})-(\ref{gd}), with the soft approximation
in Eq.~(\ref{nk}) inserted, leads to the DGLAP equation \cite{L0},
\bea
p^+\frac{d}{dp^+}\phi(x,p^+)
=\int_x^1 \frac{d\xi}{\xi}P(x/\xi,p^+)\phi(\xi,p^+)\;,
\label{con}
\eea
with the kernel
\begin{eqnarray}
P(z,p^+)=\frac{\alpha_s(p^+)}{\pi}C_F\frac{2}{(1-z)_+}\;,
\label{kgir}
\end{eqnarray}
where the variable change $\xi=x+l^+/p^+$ has been employed. The argument
of $\alpha_s$ has been chosen as the single scale $xp^+\sim (1-x)p^+$,
which is of order $p^+$. Note that the kernel $P$ differs from the
splitting function $P_{qq}=(\alpha_sC_F/\pi)(1+z^2)/(1-z)_+$ \cite{AP}
by the term $(z^2-1)/(1-z)_+$, which is finite in the $z\to 1$ limit.
The reason is that the real gluon emission is evaluated under soft
approximation as deriving $P$, while it is calculated exactly as deriving
$P_{qq}$.

\vskip 1.0cm
\centerline{\large\bf 5. Unified Logarithmic Summations}
\vskip 0.5cm

In this section we study the case, in which both the $l^+$ and $l_T$
dependences of $\phi$ in Eq.~(\ref{fsr}) are retained. It will be shown
that a unified resummation for large $x$ and a unified evolution equation
for intermediate and small $x$ are derived. Obviously, we should apply
both the Fourier and Mellin transformations to Eq.~(\ref{fsr}),
and obtain
\begin{eqnarray}
{\bar\phi}_s(N,b,p^+)=K(p^+/(N\mu),1/(b\mu),\alpha_s(\mu))\phi(N,b,p^+)\;,
\end{eqnarray}
with the soft function
\begin{eqnarray}
K&=&ig^2C_F\mu^\epsilon\int_0^1dz
\int\frac{d^{4-\epsilon}l}{(2\pi)^{4-\epsilon}}
N_{\nu\beta}(l)\frac{{\hat v}^\beta v^\nu}{v\cdot l}
\left[\frac{\delta(1-z)}{l^2}\right.
\nonumber\\
& &\left.+2\pi i\delta(l^2)\delta\left(1-z-\frac{l^+}{p^+}\right)
z^{N-1}e^{i{\bf l}_T\cdot{\bf b}}\right]-\delta K\;,
\nonumber\\
&=&\frac{\alpha_s(\mu)}{\pi}C_F\left[\ln\frac{1}{b\mu}
-K_0\left(\frac{2\nu p^+b}{N}\right)\right]\;,
\label{uk}
\end{eqnarray}
$K_0$ being the modified Bessel function. It is easy to
examine the large $b$ and $N$ limits of the above expression: for
$p^+b\gg N$, we have $K_0\to 0$ and the soft function $K$ approaches
Eq.~(\ref{kh}) for the $k_T$ resummation. For $N\gg p^+b$, we have
$K_0\approx -\ln(\nu p^+b/N)$ and $K$ approaches Eq.~(\ref{ktt}) for the
threshold resummation.

Equation (\ref{uk}) implies a characteristic scale of order
\begin{equation}
\frac{1}{b}\exp\left[-K_0\left(\frac{p^+b}{N}\right)\right]\;.
\label{uch}
\end{equation}
Following the similar procedures from
Eqs.~(\ref{dph})-(\ref{fb}), we derive the unified resummation,
\begin{equation}
\phi(N,b,p^+)=\Delta_u(N,b,p^+)\phi^{(0)},
\end{equation}
with the exponential
\begin{equation}
\Delta_u(N,b,p^+)
\exp\left[-2\int_{\exp[-K_0(p^+b/N)]/b}^{p^+}\frac{d p}{p}
\int_{\exp[-K_0(p^+b)]/b}^{p}\frac{d\mu}{\mu}
\gamma_{K}(\alpha_s(\mu))\right],
\label{fb3}
\end{equation}
which is appropriate for arbitrary $b$ and $N$. The lower bound of $p$
corresponds to Eq.~(\ref{uch}), while the lower bound of $\mu$ is
motivated by the $b\to\infty$ and $b\to 0$ limits, at which
Eq.~(\ref{fb3}) approaches Eq.~(\ref{fb}) and Eq.~(\ref{fbt}),
respectively.

In the intermediate and small $x$ regions, it is not necessary to resum the 
double logarithms $\ln^2(1/N)$. After extracting the $k_T$ resummation, the
remaining single-logarithm summation corresponds to a unification of the
DGLAP and BFKL equations, since both the $l^+$ and $l_T$ dependences have
been kept. We reexpress the function $\phi$ in the integrand of 
${\bar \phi}_{sr}$, under the Fourier transformation, as
\begin{eqnarray}
& &\phi(x+l^+/p^+,b,p^+)=\theta((1-x)p^+-l^+)\phi(x,b,p^+)
\nonumber\\
& &\hspace{1.5cm} +[\phi(x+l^+/p^+,b,p^+)-\theta((1-x)p^+-l^+)
\phi(x,b,p^+)]\;.
\label{fre}
\end{eqnarray}
The contribution from the first term is combined with ${\bar \phi}_{sv}$, 
giving the soft function $K$ for the $k_T$ resummation. The RG solution of
$K+G$ is given by 
\bea
K+G={\bar \alpha}_s(xp^+)\left[\ln(1-x)+\ln(p^+b)\right]
-\int_{1/b}^{xp^+}\frac{d\mu}{\mu}\gamma_K(\alpha_s(\mu))\;,
\label{ka}
\eea
where the first term on the right-hand side comes from the extra $\theta$
function in Eq.~(\ref{fre}). The color factor has been replaced by $N_c$,
since we are considering the gluon distribution function. The contribution
from the second term is written as
\begin{eqnarray}
& &iN_cg^2\int\frac{d^4l}{(2\pi)^4}
N_{\nu\beta}(l)\frac{{\hat v}^\beta v^\nu}{v\cdot l}
2\pi i\delta(l^2)e^{i{\bf l}_T\cdot {\bf b}}
\nonumber \\
& & \times[\phi(x+l^+/p^+,b,p^+)-\theta((1-x)p^+-l^+)\phi(x,b,p^+)]\;,
\label{fs2}
\end{eqnarray}
which will generate the splitting function below.

The master equation (\ref{meq}) then becomes 
\begin{eqnarray}
p^+\frac{d}{dp^+}\phi(x,b,p^+)&=&-2\left[\int_{1/b}^{xp^+}
\frac{d\mu}{\mu}\gamma_K(\alpha_s(\mu))
-{\bar \alpha}_s(xp^+)\ln(p^+b)\right]\phi(x,b,p^+)
\nonumber\\
& &+2{\bar\alpha}_s(xp^+)\int_x^1 dz P_{gg}(z)\phi(x/z,b,p^+)\;,
\label{ue2}
\end{eqnarray}
with the splitting function 
\begin{equation}
P_{gg}=\left[\frac{1}{(1-z)_+}+\frac{1}{z}-2+z(1-z)\right]\;,
\label{pgg}
\end{equation}
obtained from Eq.~(\ref{fs2}). The term $-2+z(1-z)$ finite as
$z\to 0$ and $z\to 1$ has been added. The term proportional to $\ln(1-x)$
in Eq.~(\ref{ka}) has been absorbed into $P_{gg}$ to arrive at the plus
distribution $1/(1-z)_+$. We first extract the exponential $\Delta$ from
the $k_T$ resummation,
\begin{eqnarray}
\Delta(x,b,Q_0,p^+)
=\exp\left(-2\int_{xQ_0}^{xp^+}\frac{dp}{p}
\left[\int_{1/b}^{p}
\frac{d\mu}{\mu}\gamma_K(\alpha_s(\mu))
-{\bar \alpha}_s(p)\ln\frac{p b}{x}\right]\right),
\label{del}
\end{eqnarray}
where $Q_0$ is an arbitrary low energy scale. It is easy to justify by
substitution that the gluon distribution function is given by 
\begin{eqnarray}
\phi(x,b,Q)&=&\Delta(x,b,Q_0,Q)\phi^{(0)}
\nonumber\\
& &+2\int_x^1 dz\int_{Q_0}^{Q}\frac{d\mu}{\mu}
{\bar\alpha}_s(x\mu)\Delta_k(x,b,\mu,Q)P_{gg}(z)\phi(x/z,b,\mu)\;,
\nonumber\\
& &
\label{nunif}
\end{eqnarray}
with $\phi^{(0)}$ the initial condition of $\phi$. Equation (\ref{nunif})
is the unified evolution equation, which can be regarded as a modified 
version of the Ciafaloni-Catani-Fiorani-Marchesini equation \cite{CCFM}.

\vskip 1.0cm
\cl{\large\bf 6. Conclusion}
\vs 0.5cm

In this paper we have demonstrated that all the known single- and
double-logarithm summations, including their unifications, can be
derived in the CS technique. The point is the treatment of the real
gluon contributions. Simply adopting soft approximations appropriate
in different kinematic regions, {\it i.e.}, neglecting the $l^+$ or $l_T$
dependence in the parton distribution function associated with the real
gluon emission, the CS technique reduces to the various logarithmic
summations. If keeping both the $l^+$ and $l_T$ dependences, a unified
resummation for large $x$ and a unified evolution equation for intermediate
and small $x$ are obtained. Our conclusion has been summarized in Table I.

The $k_T$ and threshold resummations, and the DGLAP and BFKL equations
have been widely studied and applied to many QCD processes. The unified
resummation is appropriate for the analysis of dijet production \cite{JH},
in which the transverse energy of one jet (the trigger jet) is measured,
while the other jet (the probe jet) has large rapidity up to 3.0, that
corresponds to high $x$ values. The unified evolution equation, because of
its extra $Q$ dependence at small $x$, is appropriate for the explanation
of the HERA data of the proton structure function $F_2(x,Q^2)$ \cite{H1}. 
These subjects will be discussed elsewhere.

\vskip 0.5cm
This work was supported by the National Science Council of Republic of
China under Grant No. NSC-88-2112-M-006-013.

\newpage

Table I. Logarithmic summations derived from the Collins-Soper technique 
under different soft approximations at different Bjorken variables $x$.

\vskip 0.5cm

\begin{center}
\begin{tabular}{ccccccc}
\hline
              &$|$& small $x$ &$|$& intermediate $x$ &$|$& large $x$  \\
\hline
neglect $l^+$ &$|$& BFKL equation&$|$& $k_T$ resummation &$|$&         \\ 
\hline
neglect $l_T$ &$|$&   &$|$& DGLAP equation &$|$& threshold resummation \\
\hline
no neglect    &$|$& unified      & & equation &$|$& unified resummation \\
\hline
\end{tabular}
\end{center}

\vskip 2.0cm
\centerline{\large \bf Figure Captions}
\vskip 0.5cm

\noindent
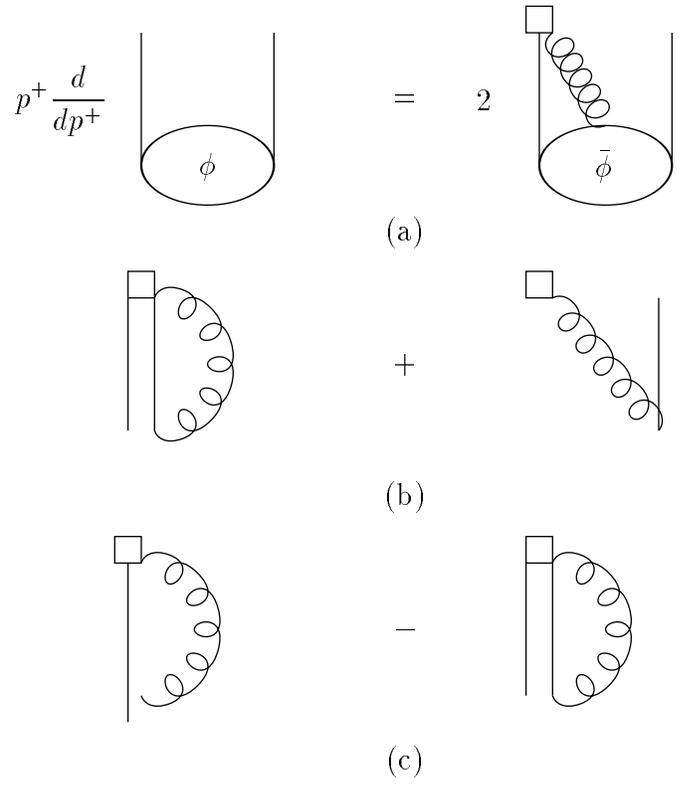
{\bf FIG. 1.} (a) The derivative $p^+d\phi/dp^+$ in the axial gauge.
(b) The soft structure and (c) the ultraviolet structure of the
$O(\alpha_s)$ subdiagram containing the special vertex.
\vskip 0.5cm

\end{document}